\documentclass[12pt]{article}
\usepackage{hyperref}
\newcommand{\be}[3]{\begin{equation}  \label{#1#2#3}}
\newcommand{\bea}[3]{\begin{eqnarray}  \label{#1#2#3}}
\newcommand{\ee}{\end{equation}}
\newcommand{\eea}{\end{eqnarray}}
\newcommand{\ba}{\begin{array}}
\newcommand{\ea}{\end{array}}

\newcommand{\resetcounter}{\setcounter{equation}{0}}  

\newcommand{\1}{{\mathbb 1}}
\newcommand{\Z}{{\mathbb Z}}
\newcommand{\N}{${\cal N}\, $}
\newcommand{\C}{\mathbb C }
\newcommand{\PP}{\mathbb P}
\newcommand{\FF}{\, \underline{F}\, }
\newcommand{\GG}{\, \underline{G}\, }
\newcommand{\HH}{\, \underline{H}\, }
\newcommand{\LL}{\,\underline{\lambda}\, }

\newcommand{\haken}{\mathbin{\hbox to 8pt{%
                 \vrule height0.4pt width7pt depth0pt
                 \kern-.4pt
                 \vrule height4pt width0.4pt depth0pt\hss}}}

\let\Large=\large
\let\large=\normalsize

\usepackage{amssymb,bbold,euscript}

\begin{document}

\begin{flushright}
\hfill{UPR-1070-T}\\
\hfill{AEI-2004-016}\\
\hfill{hep-th/0407263}

\today

\end{flushright}

\vspace{10pt}

\begin{center}{ \Large
{\bf  General \N = 1 Supersymmetric Fluxes in\\[2mm]
 Massive Type IIA String Theory
}}

\vspace{15pt}

{Klaus Behrndt$^\dagger\, ^+$ \footnote{E-mail: {\tt behrndt@aei.mpg.de}}
\quad  and \quad { Mirjam Cveti\v c}$^\dagger$ \footnote{E-mail:
{\tt cvetic@cvetic.hep.upenn.edu}}
 }
\vspace{10pt}

{\it $^\dagger$ Department of Physics and Astronomy, \\
University of Pennsylvania, Philadelphia, PA 19104-6396, USA}

\vspace{7pt}

{\it $^+$  Albert Einstein Institute,\\
 Am M\"uhlenberg 1, 14476 Golm, Germany}

\vspace{15pt}

\underline{ABSTRACT}
\end{center}

\noindent 
We study conditions on general fluxes of massive Type IIA supergravity
that lead to four-dimensional backgrounds with \N = 1 supersymmetry.
We derive these conditions in the case of SU(3)- as well as
SU(2)-structures.  SU(3)-structures imply that the internal space is
constrained to be a nearly K\"ahler manifold with all the turned on
fluxes, and the negative cosmological constant proportional to the
mass parameter, and the dilaton fixed by the quantized ratio of the
three-form and four-form fluxes. We further discuss the implications
of such flux vacua with added intersecting D6-branes, leading to the
chiral non-Abelian gauge sectors (without orientifold projections).
Examples that break SU(3)-structures to SU(2)-ones allow for the
internal space conformally flat (up to orbifold and orientifold
projections), for which we give an explicit example. These results
provide a starting point for further study of the four-dimensional
(chiral) \N = 1 supersymmetric solutions of massive Type IIA
supergravity with D-branes and fluxes, compactified on orientifolds.

\noindent

\newpage


\section{Introduction}


Insights into four-dimensional \N = 1 supersymmetric vacua of M- and
string theory with non-Abelian gauge sectors and chiral matter may
provide an important link between the M-theory and the particle
physics, describing the Standard Model and/or Grand Unified models.
Within M-theory unification the perturbative heterotic string is only
one of the ``corners'' of M-theory.  Other corners, such as Type I,
Type IIA and Type IIB superstring theory provide other, potentially
phenomenologically viable string vacua, which are related to the
heterotic ones via a web of string dualities.  In the latter framework
D-branes play an important role in constructing chiral models with
non-Abelian symmetry.

The rich structure of \N = 1 supersymmetric vacua can be significantly
increased by introducing, in addition to brane configurations, the
(supergravity) fluxes.  [There is a growing literature on the subject
of string compactifications with fluxes which was initiated in
\cite{Strominger,PS}.  A partial list of subsequent works includes,
e.g., \cite{GVW}--\cite{240},
and for recent
work, quantifying effects in terms of deformations of the original
manifold (G-structures)\cite{340}, see \cite{CGLP}--\cite{Grana04}
and
references therein.]  Typically fluxes generate a back reaction on the
original geometry of the internal space, thus changing the nature of
the internal space. The existence of Killing spinors in turn allows
for the classification of the new geometry of the internal space in
terms of specific non-trivial torsion components, which can be
classified with respect to the structure group of the internal space
(the so-called G-structures). [For a review see \cite{650} and
references therein.]

Another important effect of supergravity fluxes is the lift of the
continuous moduli space of the string vacua, i.e, fluxes introduce the
supergravity potential for the compactification moduli fields in the
effective four-dimensional theory.  The ground state solution, at the
minimum can in principle preserve supersymmetry and at the same time
fix a (sub-)set of compactification moduli. [For the intriguing new
developments in the study of flux vacua with broken supersymmetry, see
\cite{840,Louis03,Camara03,Lust04p} and references therein.]  Thus, on
one hand, the flux compactifications provide a mechanism for moduli
stabilization, one essential ingredient in the construction of
phenomenologically viable string vacua, and on the other hand, the
(probe) D-branes sectors provide the non-Abelian gauge structure and
chiral matter, another essential ingredient in reproducing the
realistic particle physics from M/string theory. Therefore the
ultimate goal of the program is to obtain consistent, explicit
constructions of (supersymmetric) flux vacua with D-brane
configurations, whose fluxes would stabilize (most/all) moduli and the
D-branes would reproduce the gauge structure and chiral matter of the
Standard/Grand-Unified models.  If such a goal were achieved, it would
provide an important link between M-theory and realistic particle
physics.

The aim of this paper is to address a specific aspect of this
program.  We focus on the systematic study of the supersymmetry
conditions for general fluxes of massive Type IIA string theory,
that yield four-dimensional \N = 1 supersymmetric vacua. This study is
part of the program that aims at shedding light on the structure of
four-dimensional \N = 1 supersymmetric vacua of Type IIA string theory
with fluxes, along with the explicit construction of D-brane
configurations.  [The past few years have seen a surge of activities
in explicit constructions of four-dimensional string vacua with
intersecting D6-branes on Type IIA orientifolds (or magnetized
D-branes on the dual Type IIB orientifolds) \cite{Blum00,AADS} with
semi-realistic particle physics \cite{Alda00,Ibanezetal,Blumetal}.
[For first \N=1 supersymmetric constructions of that type see
\cite{CSUI,CSUII}.]  Within this framework important consequences for
particle physics are due to the appearance of chiral matter at the
D6-brane intersection points in the internal space
\cite{130,Bachas}. ]

It turns out that the structure of the possible flux configurations in
Type IIA theory is very rich and the constraints are not well
understood. One should contrast this situation with that of
four-dimensional \N = 1 supersymmetric solutions of Type IIB theory
with fluxes, which is much better understood.  See, \cite{150} and
further work \cite{KachruetalI,KachruetalII} as well as recent efforts
\cite{CU,BLT} to construct supersymmetric chiral models with
(magnetized) branes and fluxes.

On the Type IIA side specific progress has nevertheless been made.  In
Ref.\ \cite{100} the \N = 1 supersymmetric vacuum given by
intersecting D6-branes in the presence of NS-NS three-form fluxes for
the massive TypeIIA supergravity was studied.  Background fluxes
induce, via back reaction on the geometry, non-zero intrinsic torsion
components or $G$-structures.  The presence of NS-NS three-form fluxes
breaks the SU(3) structures to SU(2) and the D6-branes intersect at
angles of SU(2) rotations; non-zero mass parameter corresponds to
D8-brane configurations which are orthogonal to the common cycle of
all D6-branes. The anomaly inflow indicates that the gauge theory on
intersecting (massive) D6-branes is not chiral \cite{100}.  Recent
work \cite{120} (see also \cite{350,351,110,Grana04}) that focuses on
a relationship between the constraints on fluxes of M-theory and those
of Type IIA theory provide additional insights into specific structure
of the supersymmetric flux configurations in massless Type IIA string
theory.

Most recently, in \cite{780}, further progress has been made in this
direction by analyzing the general supersymmetry constraints for
four-dimensional \N=1 supersymmetric vacua of massive Type IIA
supergravity where the Killing spinor is a SU(3) singlet, and thus the
internal manifold has SU(3) structure.  The result turned out to be
extremely constraining, with the internal geometry corresponding to
nearly-K\"ahler manifolds, with all the allowed fluxes proportional to
the mass parameter, and the dilaton determined by a ratio of the
(quantized) fluxes.  The four-dimensional negative cosmological
constant is also determined by the mass parameter and the dilaton
field, and thus in the weak string coupling limit becomes small.
There are explicit examples of nearly-K\"ahler manifolds with
supersymmetric (intersecting) three cycles that the D6-branes can
wrap; such backgrounds can therefore provide consistent string
compactifications, {\it without orientifold planes}, where the
non-Abelian chiral sector of the theory arises from the intersecting
D6-branes wrapping the supersymmetric three-cycles, while the
compactification moduli and the dilaton are fixed due to the turned on
fluxes.

In this paper we advance a number of aspects of flux compactifications
for supersymmetric vacua of massive Type IIA supergravity theory.  We
develop general techniques to study effectively the supersymmetry
conditions with the most general spinor Ansatz in the case of both,
SU(3) and SU(2) structures (Sections 2 and 3, respectively).  In
Sections 4 and 5 we explicitly derive the conditions on fluxes and
geometry for SU(3) and SU(2) structures, respectively.  In Section 4
we explicitly derive that the results of \cite{780}, provide a unique
solution with the SU(3) structure, whose internal geometry corresponds
to the nearly-K\"ahler manifold, and all the turned on fluxes and the
four-dimensional negative cosmological constant are proportional to
the mass parameter.  In Section 5 we explicitly derive a solution with
SU(2) structure where the four-dimensional space is Minkowski and the
internal space is conformally flat one. The SU(2) structure of such
solutions singles out the $T^2$ direction and we assume that all the
field and metric coefficients depend only on the $T^2$ fiber
coordinates.  This solution has very interesting implications: the
flux vacuum with the internal space conformally flat (with orbifold
and orientifold projections) allows for explicit constructions of
intersecting D6-brane configurations for which the gauge and chiral
spectrum can now be calculated explicitly, using conformal field
theory techniques.  Section 6 is further devoted to the study of
important physics implications for the vacuum solutions with the SU(3)
structure, in particular the specific examples of the nearly-K\"ahler
internal geometry with supersymmetric intersecting three-cycles that
allows for the consistent vacuum solutions with chiral non-Abelian
gauge sectors. In addition, we also highlight a possibility to address
the vacuum selection with the positive cosmological constant within
this framework.


\section{Fluxes and supersymmetry transformation in massive
 type IIA string theory}

\resetcounter


In this Section we shall spell out our notation, the relationships
between the gauge potential and field strengths and the form of
supersymmetry transformations in massive Type IIA supergravity
theory. The notation and conventions are primarily following the work
of Romans \cite{110p}; our notations are also explained in \cite{100}.

\noindent{\it Bosonic fields} \quad In massive Type IIA string theory, the
NS-NS 2-form and the RR 1-form potential combine into a gauge
invariant (massive) 2-form given by (for convention see \cite{770})
\be282
F = m B + dC_1
\ee
and the 4-form becomes
\be273
G = dC_3 + {1 \over 2\, m} F \wedge F
\ee
were $C$ is the RR 3-form potential.  Due to the Chern-Simons terms,
both forms are not closed but
\be732 
dF = m H \quad , \qquad dG =  F \wedge H \ .
\ee 
For the sake of notational simplicity we suppressed a subscript
indicating the $n$-index of a specific $n$-form.  In the massless
limit ($m=0$) one gets
\[
F^{(0)} = dC_1 \qquad ,  \quad 
G = dC_3 +  F^{(0)} \wedge B = 
d(C_3 +  \, C_1\wedge B)  +  C_1 \wedge H
\]
Note, only in the massless case the 2-form $F=F^{(0)}= dC_1$ is exact.

\noindent{\it Supersymmetry transformations}\quad Unbroken
supersymmetry requires the existence of at least one Killing spinor
$\epsilon$, which is fixed by the vanishing of the fermionic
supersymmetry transformations [for a purely bosonic configuration, the
variations of the bosons vanish trivially]. These variations have been
first setup for massive type IIA supergravity in Einstein frame
\cite{110p}, but we will use the string frame and the fermionic
variations become \cite{770}
\be040
\ba{rcl}
\delta \psi_M &=& \Big\{D_M  + {1 \over 8} H_M \Gamma_{11}
+ {1 \over 8} \, e^{\phi} \Big[ \, m  \,  \Gamma_M  +
  F \Gamma_{M}  \, \Gamma_{11}
+  \, G \, \Gamma_M  \Big] \Big\} \epsilon \ ,
\\[3mm]
\delta \lambda &=&  \Big\{ \partial \phi 
      + {1 \over 12} \, H \, \Gamma_{11} 
       + {1 \over 4} e^{\phi}\Big[ 5\, m + 3 F \, \Gamma_{11}
      +  G \Big] \Big\}  \epsilon \ .
\ea
\ee
and we used the abbreviations
\be050
\partial \equiv \Gamma^M \partial_M \  , \quad
H = H_{PQR} \Gamma^{PQR} \  , \quad  H_M = H_{MPQ} \Gamma^{PQ} \ ,
\quad {\rm etc.}
\ee
Apart from the differential forms that we introduced already, the mass
parameter is denoted by $m$ and $\phi$ is the dilaton.  In type IIA
supergravity, the Killing spinor $\epsilon$ is Majorana and can be
decomposed into two Majorana-Weyl spinors of opposite chirality. The
massless case can be lifted to M-theory and this Majorana spinor
becomes the 11-dimensional Killing spinor.  We will come back to our
spinor convention below.

\noindent{\it Ansatz for the Metric and field strength}\quad We are
interested in compactifications on to a 4-d spacetime that is either
flat or anti deSitter, i.e. up to warping the 10-d space time
factorizes $M_{10} = X_{1,3} \times Y_6$ and we write the metric
Ansatz as
\be060 
ds^2 = e^{2A(y)} \, \Big[ g_{\mu\nu} dx^\mu dx^\nu  + 
 h_{mn}(y) \, dy^m dy^n \, \Big] \ .
\ee
where $g_{\mu\nu}$ is either flat or $AdS_4$ and $h_{mn}$ is the
metric on $Y_6$ and the warp factors depend only on the coordinates
of the internal space. An especially interesting question, which we
will address in more detail below, are the constraints that allow for
a flat internal/external space. Consistent with this metric Ansatz is
the assumption that the fluxes associated with the forms $F$ and $H$
have non-zero components only in the internal space $Y_6$ whereas $G$
may have in addition a Freud-Rubin parameter $\lambda$:
\be062
\ba{l}
F = {1 \over 2} \, F_{mn} dy^m \wedge dy^n \quad , \qquad
H = {1 \over 3} \, H_{mnp} dy^m \wedge dy^n \wedge dy^p \ , \\[2mm]
G = \lambda \, dx^0 \wedge dx^1 \wedge dx^2 \wedge dx^3
+ {1 \over 4}\, G_{mnpq} \, dy^m \wedge dy^n \wedge dy^p \wedge dy^q \ .
\ea
\ee
Note, all forms as well as the warp factor and the dilaton are in
general functions of coordinates $y^m$ of the the internal space.
With these Ans\"atze the gravitino variation splits into an external
and internal part and with $D_M = \nabla_M + {1 \over 2} \Gamma_M{}^N
\partial_N A$, we find for the variations in (\ref{040})
\be668
\ba{l}
0= \Big[ \, {1 \over 2} \, e^A \, \Gamma^\mu \nabla_\mu + 
\partial A 
+ {1 \over 4} e^{\phi +A} \Big( m +  \FF 
\Gamma_{11} + \GG - 4 \LL \Gamma_{0123}
\Big)\Big]\, \epsilon \ , \\[3mm]
0=\Big[\, \nabla_m  + {1 \over 8}  \HH_m \Gamma_{11} 
- {1 \over 4}\, e^A  \Gamma_m (\Gamma^\mu \nabla_\mu) \\
\qquad \qquad -  {1 \over 2} e^{\phi +A} 
\Big( \FF_m \Gamma_{11} + 2 
\GG_m - 2 \LL \Gamma_{0123} \Gamma_m
 \Big) \Big] \epsilon \ , \\[3mm]
0= \Big[\partial \phi + {1 \over 12}  \HH  \Gamma_{11} 
+ {1\over 4} e^{\phi +A} \Big( 5 m + 3 \, 
\FF \Gamma_{11} +   \GG + 4 \LL \Gamma_{0123}
 \Big) \Big]\, \epsilon
\ea
\ee
where $(\nabla_\mu , \nabla_m)$ are the covariant derivatives with
respect to the metrics ($g_{\mu\nu} , h_{m,n}$) and we defined
the rescaled forms
\be023
\underline{F} \equiv e^{-2A} F \quad , \qquad
\underline G \equiv e^{-4A} G \quad , \qquad
\underline H \equiv e^{-2A} H\quad , \qquad
\LL \equiv e^{-4A} \, \lambda
\ee
In order to proceed, we decompose the $\Gamma$-matrices as usual
\be070
\ba{l}
\Gamma^{\mu} =  \hat \gamma^{\mu} \otimes {\mathbb 1}
\quad, \qquad
\Gamma^{m+3} = \hat \gamma^5 \otimes \gamma^m
\quad , \qquad
\Gamma^{11} = - \hat \gamma^5 \otimes \gamma^7 \ , \\[2mm]
\hat \gamma^5 = i \hat \gamma^0 \hat \gamma^1 \hat \gamma^2 \hat \gamma^3
    \quad ,\qquad
\gamma^7 = i \gamma^1 \gamma^2 \gamma^3 \gamma^4 \gamma^5 \gamma^6 \ .
\ea \ee
We use the Majorana representation so that $\Gamma^{11}$, $\hat
\gamma^\mu$ are real and $\hat \gamma^5$, $\gamma^7$ and $\gamma^m$
are imaginary and anti-symmetric and to avoid confusions, we hatted
the 4-dimensional $\gamma$-matrices and use $m,n, \ldots$ for internal
indices and Greek for external indices. In addition, the spinor has to
be decomposed into an 4-d spinor and an internal spinor, which will
discuss in more detail in the next section.  If the external space is
flat, the 4-d spinor is covariantly constant, but if want to include
$AdS_4$ vacua, which appear naturally as supersymmetric vacua, the
covariant derivative of the 4-d spinor gives essentially the
superpotential. If we denote the 4-d Killing spinor with $\theta$ and
if it is Weyl, we introduce the complex superpotential $W$ by
\be399
\nabla_\mu \theta = \hat \gamma_\mu \bar W \theta^\star \ .
\ee
As integrability constraint, one finds that the 4-d space time is
anti-deSitter with $\Lambda = - |W|^2$ as the 4-d (negative)
cosmological constant.


\section{Killing spinors and $G$-structures}

\resetcounter


As for the bosonic fields, also the 10-d spinors have to be decomposed
into an external and internal part and the general 10-d Killing spinor
can be expanded in all independent internal and external spinors so
that we can write in general
\be542
\epsilon = \theta_i \otimes \eta_i + cc
\ee
where $\theta_i$ and $\eta_i$ are the four- and six-dimensional
spinors, respectively.  Recall, we use the convention that $\epsilon$
is a general Majorana spinor, not necessarily Weyl, and
this spinor is now expanded in all independent internal and external
spinors, where all internal spinors are chosen to be Weyl and the
external are in general Dirac. Since these spinors have to be singlets
under the structure group $G$, this most general case highly restricts
the geometry of the internal space; e.g.\ only the flat space or the
spheres can support four independent internal Weyl spinors.  Note, in
11 dimensions the Majorana Killing spinor $\epsilon$ is expanded in up
to eight internal and external Majorana spinors, which in the 10
dimensional spinor Ansatz are combined into internal Weyl and external
Dirac spinors.

Since the spinors are singlets under this group, the resulting
differential forms constructed as fermionic bi-linears are singlets
under the structure group and define $G$-structures.  We are
interested in the two cases: $(i)$ of SU(3) structures and $(ii)$ of
SU(2) structures.  In the first case, only one singlet spinor $\eta$
can appear, which implies for our Ansatz that: $\eta_i = a_i \eta$,
with some complex coefficients $a_i$. The resulting 4-d spinor:
$\theta = a_i \theta_i$ is one Dirac spinor, which can be written in
terms of two Weyl or two Majorana spinors. If there are no further
constraints on these two spinors, we obtain an \N=2 , $D$=4 vacuum.
Note, if there are no fluxes, the structure group determines the
holonomy of the space and as expected an SU(3) holonomy space leaves
eight supercharges unbroken. For the SU(2) case we can define two
singlet spinors on the internal space, which can be combined into one
doublet and also the 4-d spinors can combined into one doublet of two
Dirac spinors.  With no fluxes, the holonomy is SU(2) and hence the
space factorize into $Y_6 = T_2 \times X_4$, where $T_2$ is flat and
$X_4$ has SU(2) holonomy, i.e.\ in the compact case it is K3 and the
vacuum has \N=4 supersymmetry.  We are of course interested in \N=1
vacua and therefore we have to break further supersymmetry by imposing
a constraint on the 4-d Dirac spinors, which can be understood as a
flux-induced (partial) supersymmetry breaking. Actually, non-zero
fluxes correspond to specific torsion components and the corresponding
relations between fluxes and torsion will be derived in the next
section. But let us now discuss in more detail the SU(3) and SU(2)
case.

\subsection*{SU(3) structure}

\noindent
In this case, we have a single internal spinor $\eta$ and the external
spinor can be any Dirac spinor, which can be decomposed into two Weyl
spinors of opposite chirality. Therefore, the spinor Ansatz becomes
\[
\epsilon =  c_i \theta_i \otimes \eta + cc 
\]
with the complex coefficients $c_i$ ($i=1,2$) and $\hat \gamma^5
\theta_i = (\sigma_3 \eta)_i$ where $\sigma_3 $ is third Pauli
matrix\footnote{With the usual definition: $\sigma_1 = \Big(\ba{cc} &
1 \\[-1mm] 1 & \ea \Big) \ , \quad \sigma_2 = \Big(\ba{cc} & -i
\\[-1mm] i & \ea \Big) \ , \quad \sigma_3 = \Big(\ba{cc} 1 &\\[-1mm] &
-1 \ea \Big)\ .  $ }.  In our notation the complex conjugate spinors
$\theta_i^\star$ have opposite chiralities and hence the truncation to
an \N=1 vacuum is done by
\[
\theta_i^\star = (\sigma_1 \theta)_i 
\]
 which is
consistent with the Weyl property. Setting $\theta_1 = \theta$, the
spinor Ansatz for an \N=1 vacuum becomes
\be072
\epsilon =  (a \theta + b \theta^\star) \otimes \eta + cc 
\ = \ \theta \otimes ( a \eta + b^\star \eta^\star ) + cc \ .
\ee
There are two special cases: if $ab = 0$ the 4-d spinor is Weyl
and if $b = a^\star$ we have a Majorana spinor.

On the other hand, the SU(3) singlet spinor is Weyl and we choose
\be138
\eta = {1 \over \sqrt 2}  
\, ({\mathbb 1} + \gamma^7) \, \eta_0 \ .
\ee
with $\eta_0$ being a constant spinor. Being an SU(3) singlet spinor,
$\eta$ satisfies the projectors
\be230
\ba{rcl}
(\gamma_m  + i J_{mn} \gamma^n ) \, \eta &=& 0 \\[2mm]
(\gamma_{mn} -  i\, J_{mn}) \, \eta &=& {i \over 2} \,
\Omega_{mnp} \gamma^p  \, \eta^\star \ ,
 \\[2mm]
(\gamma_{mnp} - 3 i J_{[mn} \gamma_{p]} ) \, \eta &=& i \,
\Omega_{mnp} \eta^\star\ .
\ea
\ee
where the complex structure and the holomorphic 3-form
are introduced by
\be150
\eta^\dagger \, \gamma_{mn} \eta = i \, J_{mn}
\quad , \qquad
\eta \gamma_{mnp} \eta = i \, \Omega_{mnp} 
\ee
[with $1 = \eta^\dagger \eta$]. Note, these are the only differential
forms that can be constructed from a single chiral spinor and for
non-zero fluxes they are {\em not} covariantly constant nor closed and
this failure is related to non-vanishing intrinsic torsion components.
Following the literature \cite{630, Gauntlettetal,
Louisetal,LustetalII}, one introduces five classes ${\cal W}^i$ by
\be526
\ba{rcl}
dJ &=& {3 i \over 4}\,  ( {\cal W}_1 \bar \Omega -
 \bar{\cal W}_1 \Omega ) +  {\cal W}_3 + J \wedge  {\cal W}_4 \ ,\\[2mm]
d\Omega &=&   {\cal W}_1 J \wedge J + J \wedge  {\cal W}_2
+ \Omega \wedge  {\cal W}_5
\ea
\ee
with the constraints: $J \wedge J \wedge {\cal W}_2 =J \wedge {\cal
W}_3 = \Omega \wedge {\cal W}_3=0 $.  Depending on which torsion
components are non-zero, one can classify the geometry of the internal
space. E.g., if only ${\cal W}_1 \neq 0$ the space is called nearly
K\"ahler, for ${\cal W}_2 \neq 0$ almost K\"ahler, the space is
complex if ${\cal W}_1 ={\cal W}_2= 0$ and it is K\"ahler if only
${\cal W}_5 \neq 0$. 


\bigskip

\noindent
{\em SU(3) decomposition of fluxes}


\noindent
This decomposition is done by employing the holomorphic projector: ${1
\over 2} (\1 \pm i J)$ to distinguish between holomorphic and
anti-holomorphic indices. We will indicate this by the labels at the
forms, but let us stress that this makes sense only locally.  Since
the internal space is in general not a complex manifold, one cannot
introduce global holomorphic forms.

On the 6-dimensional internal space $Y_6$, the 4-form $G$ and the
2-form $F$ have 15 and the 3-form 20 components, which decomposes
as follows
\be883
[G] = {\bf 8 + 1 + 3 + \bar 3} = 
[G^{(2,2)}] + [G^{(2,2)}_0] + [G^{(3,1)}] + [G^{(1,3)}]
\ee
with the singlet $G^{(2,2)}_0 = G_{mnpq} J^{mn} J^{pq}$ and the
vector representations\footnote{Here and in the following we use the
convention: $(\Omega \haken G)_m \equiv \Omega_{abc} G^{abc}{}_m$.}:
$G^{(3,1)} = \Omega \haken G$ and its complex conjugate: $G^{(1,3)} =
\bar \Omega \haken G$; the remaining components comprise 
the (2,2)-forms obeying: $G^{(2,2)}\wedge J =0$.

Similarly, the components of the 2-form $F$ decomposes as
\be283
[F] = {\bf 8 + 1 + 3 + \bar 3} = 
[F^{(1,1)}] + [F^{(1,1)}_0] +[F^{(2,0)}] + [F^{(0,2)}]
\ee
with the singlet: $F^{(1,1)}_0 = J\haken F$, the vectors $F^{(2,0)} =
F \haken \Omega$ and its complex conjugate. The remaining components
represent an adjoint of SU(3) satisfy: $F^{(1,1)} \wedge J\wedge
J = 0$. 

Finally, the 3-form decomposes as
\be872
\ba{rcl}
[H] &=& {\bf 6 + \bar 6 + 3 + \bar 3 + 1 + \bar 1} \\[2mm]
 &=& [H^{(2,1)}] +
[H^{(1,2)}]+ [H^{(2,1)}_0] +[H^{(1,2)}_0] + [H^{(3,0)}] +[H^{(0,3)}]
\ea
\ee
where the singlets are now: $H^{(3,0)} = \Omega \haken H$ and its
complex conjugate; the two vectors $H^{(2,1)}_0$ and $H^{(1,2)}_0$ are
the two holomorphic projections of the (real) vector: $J \haken H$
and the ${\bf 6 + \bar 6}$ are the primitive (2,1) and (1,2) forms
fulfilling $H \wedge J = 0$.


\subsection*{SU(2) structure}


If the structure is broken to SU(2), one finds two singlet spinors
$\eta_k$ and we choose them of opposite chirality, i.e.
\be522
\gamma^7  \eta_k = (\sigma_3 \eta)_k  
\ee
with: $(\sigma_3 \eta)_k \equiv (\sigma_3)_k{}^l \, \eta_l$. 
In addition there two external Dirac spinors $\theta_i$
and in order to obtain an \N=1 vacuum we have to impose
projectors. As for the SU(3) case, we can first truncate each Dirac
spinor into one Weyl spinor and write as in (\ref{072})
\be331
\epsilon =  (a_1 \theta_1 + b_1 \theta^\star_1) \otimes \eta_1 
+ (a_2 \theta_2 + b_2 \theta^\star_2) \otimes \eta_2 
+ cc 
\ee
where $a_i$, $b_i$ are complex coefficients. Now, the truncation to
\N=1 is given by the following projectors obeyed by the 4-d spinors
\[
\theta_i = \hat \gamma^5 \theta_i \quad , \qquad
\theta_i = (\sigma_1 \theta)_i 
\]
so that both spinors have the same chirality and moreover $\theta_2
=\theta_1 \equiv \theta$ [$\sigma_1$ can also be replaced by
$\sigma_2$].  Thus, $\epsilon$ becomes
\be729
\epsilon = \theta \otimes (a_i \eta_i + b_i^\star \eta_i^\star) \, + \, cc
\ =\ (\theta a_i + \theta^\star b_i) \otimes \eta_i + cc \ .
\ee
Here, $\theta$ is a single Weyl spinor, but if we set in (\ref{331}):
$a_1=b_1=c_1$ and $a_2 = - b_2 = i\, c_2$, this spinor becomes:
$\epsilon = c_1 \theta^M_1 \otimes \eta_1 + i c_2 \hat \gamma^5
\theta^M_2 \otimes \eta_2 + cc$, with the 4-d Majorana spinor
$\theta^M_i = \theta_i + \theta^\star_i$.  For this Majorana case, the
truncation to \N=1 vacuum corresponds to the relations $i\hat \gamma^5
\theta^M_1 = \theta^M_2$ between the two Majorana spinors $\theta^M_1$
and $\theta^M_2$.

The internal spinors $\eta_i$ have opposite chirality and one of
them can be taken as the SU(3) singlet, say $\eta_1$. The other can then
be introduced by means of a vector field $v_m$
\[
\eta_2 = v \, \eta_1 \equiv v_m \gamma^m \, \eta_1 \qquad {\rm with:}
\quad |v| = 1 \ .
\]
With these spinors, we define the following differential forms
\be723
\Lambda^{(n)}_{kl} =  \eta^\dagger_k \gamma^{(n)} \eta_l
\quad , \qquad 
\Sigma^{(n)}_{kl} =  \eta_k^T \gamma^{(n)} \eta_l
\ee
where $\gamma^{(n)} \equiv \gamma_{m_1 m_2 \cdots m_n}$.  Each of this
form is a $2 \times 2$ matrix and in the following we will use a
matrix notation in terms of  Pauli matrices. The spinors are
normalized and chiral so that
\[
\Lambda^{(0)} = \1 \quad , \qquad \Sigma^{(0)} = 0 
\]
We recover the SU(3) expressions in (\ref{150}) by setting $v=0$
\[
{\rm SU(3)\ case:} \qquad \Lambda^{(2)} = i \, J \ , \quad
\Lambda^{(4)} = - {1 \over 2} J\wedge J \ , \quad \Sigma^{(3)} = i \,
\Omega
\]
which does not allow for a vector. But in the SU(2) case one can
construct two vectors, which can be combined into one complex vector:
$ \eta_1^\dagger \gamma_m \eta_2 = v_m + i J_m{}^n v_n \equiv 2
v_m^{(+)}$. With the definition
\[
u^m = J^m{}_n v^n
\]
we find the following useful relations ($v \equiv v_m \gamma^m$ etc.)
\be947
\eta_k = v \,  (\sigma_1)_k{}^l \, \eta_l =
 u \, (\sigma_2)_k{}^l \, \eta_l  \qquad {\rm or:}
\qquad  (v - i u) \, \eta_1 = (v + i u)\, \eta_2 = 0 
\ee
and hence we get for the 1-forms
\[
\Lambda^{(1)} = v \sigma_1 - u \sigma_2 \quad ,
\qquad
\Sigma^{(1)} = 0 \ .
\]
With this 1-form, the spinors $\eta_k$ satisfy the projectors
\[
(\1 + {1 \over 2} \Lambda_m \gamma^m) \eta^\star = 0
\quad , \quad \Lambda_m \gamma^m \eta =0 \ .
\]
The existence of the (complex) vector implies, that the 6-d internal
space is locally a complex line bundle over a 4-d base space.  We find
for the 2-forms
\be223
\Lambda^{(2)} = i \1 \, J^{(0)} +i \, \sigma_3 \, v \wedge u
\quad , \qquad
\Sigma^{(2)} =  \sigma_2 \, \hat\Omega \ .
\ee
Similar to the SU(3) case, $J^{(0)}$ is an almost complex structure on
the base and can be used to project onto holomorphic and
anti-holomorphic components. In addition, $\hat \Omega_{ab} \equiv
\Omega_{abc} v^c$ is a (2,0)-form and its complex conjugate is
(0,2)-form.  If the spinors would be covariantly constant, as in the
absence of fluxes, these forms would identify the base as an
hyper-K\"ahler space. Note, we identified the symplectic 2-form by the
relation to the SU(3) case, but actually all three 2-forms are on
equal footing and one can also pick another one.  The two spinors
satisfy moreover the relations
\be541
\hat \Omega^{ab} \gamma_{ab} \eta^\star = - 8 \, \sigma_2 \eta
\quad ,\qquad
\hat \Omega^{ab} \gamma_{ab} \eta = 0
\quad , \qquad
J^{(0)}_{ab} \gamma^{ab} \eta = 4 i \, \eta
\ee
[recall, we are using a matrix notation and we have the doublet spinor
$\eta \equiv (\eta_1 , \eta_2)$].  There is no 3-form on the base and
hence all 3-forms have to have one leg in the fiber and we find
\be752
\ba{rcl}
\Lambda^{(3)} &=& \Lambda^{(1)} \wedge \Lambda^{(2)} = 
i (\sigma_1 v - \sigma_2 u ) \wedge  J^{(0)} 
\  , \\[2mm]
\Sigma^{(3)} &=& -\Lambda^{(1)} \wedge \Sigma^{(2)} = 
-  (\1 u - i \sigma_3 v )\wedge \hat \Omega \ .
\ea
\ee
The 4- and 5-forms are dual to the 2- and 1-forms. where one has
however to take into account the multiplication with $\sigma_2$.  For
the 4-forms one finds
\be884
\Lambda^{(4)} = - \1 \, J^{(0)} \wedge J^{(0)}
+i\, \sigma_3 u \wedge v \wedge J^{(0)} 
\quad  , \qquad
\Sigma^{(4)} =  \, \sigma_1 u \wedge v \wedge \hat \Omega_{ab}  
\ee
and the 5-forms read
\[
\Lambda^{(5)} = \Lambda^{(1)} \wedge \Lambda^{(4)} = 
(v \sigma_1 - u \sigma_2 )\wedge \Lambda^{(4)}
\quad , \qquad
\Sigma^{(5)} = 0
\]
As in the SU(3) case, these forms are not closed if fluxes are
present, which again is related to non-vanishing torsion components.

For the SU(3) case the geometry was described by the almost complex
structure $J$ and the holomorphic 3-form $\Omega$.  Here in the SU(2)
case, the 6-d geometry is fixed by the triplet ($v, J^{(0)} , \hat
\Omega^{(2,0)}$). If all fluxes vanish, the SU(3) case corresponds to
a Calabi-Yau space whereas the SU(2) case yields a factorization of
the internal space into $T^2 \times K3$ (if we assume compactness).
As we will see below, the SU(2) case allows for much more general
fluxes than the highly constrained SU(3) case.


\subsection*{Summary}


When truncated to an \N=1 vacuum, both spinor Ans\"atze can
be written as
\[
\epsilon = \theta \otimes \chi + \theta^\star \otimes \chi^\star
\]
where $\chi$ is not chiral spinor but: $\chi = a \eta + b^\star
\eta^\star$ for the SU(3) case and: $\chi = a_i \eta_i + b_i^\star
\eta_i^\star$ for the SU(2) case. If we introduce the superpotential
as in (\ref{399}) we find
\[
\Gamma^\mu \nabla_\mu \epsilon = - 4 W \, \theta \otimes \chi^\star 
+ 4 \bar W \, \theta^\star \otimes \chi \ .
\]
The 4-dimensional spinor $\theta$ is chiral and therefore all terms
coming with $\theta$ and $\theta^\star$ have to vanish separately.  If
we collect all terms ${\cal O}(\theta)$,  the variations
(\ref{668}) become
\bea761
-\partial A \chi + 2\, e^A \, W \chi^\star &=& {1 \over 4} e^{\phi + A}
\Big[ m - \FF \, \gamma_7  +
\GG + 4 i \LL \Big] \chi \ , \\[3mm] \label{329}
- \Big[ \partial \phi  - {1 \over 12}  \HH  \gamma^7\Big] \chi
&=&
{1 \over 4} e^{\phi +A} \Big[ 5m - 3 \FF  \gamma^7
+ \GG - 4 i \LL \Big] \chi \, , \\[3mm] \label{324}
\Big[\nabla_m + {1 \over 8}  \HH_m  \gamma^7 \Big] \chi
- e^A\,  W \, \gamma_m  \chi^\star &=& -e^{\phi + A}
\Big[ \, {1 \over 2} \FF_m \, \gamma^7 -   \GG_m - i \LL 
 \gamma_m \Big] \chi \, .
\eea
The complex conjugate of these equations yield the terms coming together
with $\theta^\star$.


\section{Fluxes and geometry for SU(3) structures}

\resetcounter


Now we have all the tools to explore in detail the Killing spinor
equations. With SU(3) structures, the internal spinor
reads
\[
\chi = a \eta + b^\star \eta^\star \ .
\]
A special case is if $\chi$ is chiral, i.e.\ $a$ or $b$ is zero and the
10-dimensional spinor $\epsilon$ is Majorana-Weyl. Another special
case would be $b^\star = a$, where the 4-d spinor is Majorana.
Let us discuss the different cases separately.


\subsection*{Weyl case}


Setting $b=0$ (or $a=0$), the lhs and rhs of the equations (\ref{761})
-- (\ref{324}) have opposite chirality and have to vanish separately!
One finds the same constraints as the one derived for $m=0$ in
\cite{110, 120}, namely: $F= G = H^{(3,0)} = 0$
\footnote{Since $H$ is a real form, this implies also that
$H^{(0,3)}=0$.} and $d \phi = H \haken J^2 \equiv H \haken (J \wedge
J)$.  The vanishing of $F$ and $G$ follows from the internal variation
(\ref{324}), which yields after using the expression in (\ref{230})
\be328
G^{(3,1)} = 0 \quad , \qquad  \FF_m^{\ l}J_{nl} + 6 
      \GG_{mnpq} J^{pq} + 
     4  \LL \delta_{mn} = 0 \ .
\ee
By taking the trace, the singlet $F^{(1,1)}_0$ is canceled by
$\LL$ and all other components have to vanish
\[
F \sim \lambda J \quad , \qquad G_{mnpq} = 0 .
\]
Note, in real notation, $F^{(1,1)}$ satisfies: $(1 \pm iJ) \cdot
F^{(1,1)} \cdot (\1 \mp i J) = 0$ yielding that $F^{(1,1)} \cdot J$ is
a symmetric matrix. {From} the rhs of the external variation
(\ref{761}), we find a complex constraints which set the mass to zero
and another constraint on the Freud-Rubin parameter $\LL$ and
$F^{(1,1)}_0$ which is in contradiction with the relation derived from
the internal variation. Therefore one has to infer: $\LL = \lambda=
0$, $m = 0$.  Contracting the lhs of (\ref{761}) with $\eta^T$ and
$\eta^\dagger$ one gets moreover that: $W=0$ and hence: $dA = 0$. So,
all RR fields have to be zero, but the NS-$H$-field can still be
non-zero. Setting $A=0$, we find {from} the dilatino variation (\ref{329}) 
\be928
 H \haken J^2 =  d \phi
\qquad , \quad H\haken \Omega =0 
\ee
i.e.\ the dilaton is fixed by $H^{(2,1)}_0$.  These $H$-field
components enter the differential equation fixing the Killing spinor
and by investigating the torsion classes one finds that the internal
space is in general non-K\"ahlerian \cite{630, LustetalII}. 

In summary, if the 10-d spinor is Majorana-Weyl or equivalently if $ab
=0$ in our spinor, the mass and all RR-fields have to vanish:
\be772
F_{mn} = G_{mnpq} = W = dA = m=0
\ee
and only the fields from the NS-sector can be non-trivial.  The
holomorphic part $H^{(3,0)}$ has to vanish, $H^{(2,1)}_0$ fixes the
dilaton and the internal
space is non-K\"ahlerian. One might have expected this result, because
the NS-sector is common to all string models, and a common solution
can only be described by one 10-d Majorana-Weyl (Killing) spinor.  An
explicit example that solves these equations is the NS5-brane
supergravity solution, but there are also other examples
\cite{Strominger, LustetalII}.


\subsection*{Majorana case}


Another interesting case is given by $a = b^\star$, where
4-dimensional spinor $a \theta + b \theta^\star$ in (\ref{729}) is
Majorana. This implies that the 10-d spinor cannot be Weyl and hence
corresponds to the more generic situation for type IIA models.
In this case
\[
\chi = a (\eta + \eta^\star)
\]
and we can separate the real and imaginary part of the equations
(\ref{761}) -- (\ref{324}), which will give us two sets of
equations. This case has been discussed already in \cite{780}, but
since our notation here differ, let us summarize this
case.

If we define $a = r e^{i\alpha}$ and $W = e^{2i \alpha} (W_1 + i W_2)$
with $W_1$ and $W_2$ real, the first set becomes
\bea229
e^A\, W_1 \eta &=&
{1 \over 8} e^{\phi +A}  (m +   \GG)\eta     \\ \label{361}
-  \HH  \eta^\star&=& 3\, e^{\phi+A} (5 m + \GG) \eta   \\  \label{731}
i (\partial_m \alpha) \eta^\star + {1\over 8} \HH_m \, \eta^\star
+    e^A\, W_1 \gamma_m \eta &=&
e^{\phi +A} \GG_m  \eta 
\eea
With the formulae in (\ref{230}), the first two equations give:
$G^{(1,3)}=0$ and $J \haken H=0$ (i.e.\ $H \wedge J=0$); in the last
equation the singlets give the constraint
\be977
d \alpha = 0 \qquad , \quad
6 \, W_1 - 3 \, H_0 =
e^{\phi} \, G_0 
\ee
with
\be774
G = G_0 \, J \wedge J \ , \ H = H_0 \, {\rm Im} \Omega
\ee
which are the singlet components under an SU(3) decomposition. 
With (\ref{229}) and (\ref{361}) we find for these singlets
\be671
W_1 = -{m \over 10} \ , \qquad G_0 = {m \over 20} \ , \qquad
H_0 = -{2m \over 5} \, e^{\phi}  \ .
\ee  
The remaining components in $G_m$ and $H_m$ in (\ref{731}) cannot
cancel, due to the different holomorphic structure and therefore have
to vanish.  Note, all components of $G$ and $H$ can be obtained from
$\hat G_{mn} \equiv G_{mnpq}J^{pq}$ and $\hat H_{mn} \equiv H_{mpq}
\Omega^{pq}{}_n$ by different chiral projections. E.g.\ the (1,1) part
in $\hat H_{mn}$ are the ${\bf 1 + \bar 1}$ (ie.\ $H^{(3,0)}$ and
$H^{(0,3)}$) and the (2,0) + (0,2) components are in total $3 \times
3=9$ complex or ${\bf 18 = 6 + 3 + \bar 6 + \bar 3}$ real components;
see also (\ref{872}).

The remaining equations obtained from (\ref{761}) -- (\ref{324}) are
\bea222
    -  \partial A \eta + 2i \, e^{A} W_2 \eta^\star &=&
{1 \over 4} e^{\phi +A} 
\, (  \FF + 4i  \LL ) \eta^\star 
  \\ \label{629}
  - \partial \phi \, \eta &=&
      {1 \over 4} e^{\phi + A} ( \, 3 \FF - 4 i \LL ) 
     \, \eta^\star  \\  
 ( \hat \nabla_m  +\partial_m r ) \eta + i  \gamma_m \, e^A W_2 
\eta^\star &=&
 {1 \over 2} e^{\phi +A} ( \FF_m + 2i  \LL \gamma_m\, ) \eta^\star
\eea
These are differential equations for the warp factors,
the dilaton and additional constraints on the $W_{1/2}$, $\lambda$ as
well as $F^{(1,1)}_0$.  One finds
\be882
\textstyle{
F_0 = - {2 \over 9} \, \lambda \, e^{-4A} 
\ , \ W_2 = - {2 \over 3} \lambda\, e^{\phi - 3A}
} \ ,
\ee
and moreover
\be661
\ba{rcl}
 \nabla_m  \hat \eta &=&  {1 \over 2}
e^{\phi +A} (F_{mn} \gamma^n\, + i \, {32 \over 9} \LL \gamma_m) 
\hat \eta^\star \ ,
\\[1mm] \label{663}
\partial_m A &=& -{i \over 8} e^{\phi +A} \, \Omega_{mpq} \FF^{pq}
\ea
\ee
with $\hat \eta = e^r \eta$.

But it is not enough to consider the supersymmetry variation, one has
also to ensure that the 10-d equations of motion for $G$ and $H$ are
solved as well as the Bianchi identities.  The only solution that we
found requires \cite{780}
\be557
d\phi = dA = 0 \ , \quad F= F_0 J
\ee
with constant $G_0$, $H_0$ given by (\ref{671}) and $F_0$ by
(\ref{882}).  Note, $A$ being constant, they can be scaled
away and we can set $A=0$ from the very beginning. Thus, eq.\
(\ref{671}) implies that the dilaton is fixed by the ratio of the
(quantized) fluxes
\be837
\textstyle{
e^\phi = - { H_0 \over 8\,  G_0} \ .
}
\ee
The differential equation for the spinor becomes finally
\be823
\textstyle{
\nabla_m \eta = i \,  { 17 \over 9} \, e^\phi \lambda \gamma_m \eta^\star
} 
\ee
which identifies the internal space as a nearly K\"ahler manifold,
which is Einstein but neither complex nor K\"ahler! 
It can also be written as
\[
\textstyle{
0=\hat \nabla_m \eta = \Big[  \nabla_m + {34 \over 9} e^\phi
\, {\rm Re} \, \Omega_{mpq} \gamma^{pq}  \Big] \eta  
}
\]
and hence the spinor is covariantly constant with respect to the Bismut
connection (its holonomy is in SU(3)). It is straightforward to verify
that in this case only ${\cal W}_1 \neq 0$ whereas all other torsion
classes in (\ref{526}) vanish. In fact, $dJ = {17 \over 18} \lambda \,
e^\phi {\rm Im} \Omega$, $d\, {\rm Re} \Omega \sim J \wedge J$ which
ensures $dF = m H$, $dH = 0$ and $d\, {^\star H} \sim G \wedge
G$. This also fixes the Freud-Rubin parameter in terms of the mass:
\[
\textstyle{
\sqrt{85\over 2} \, \lambda  = 9 m \ .
}
\]
In the limit of vanishing mass, our solution becomes trivial, i.e.\
all fluxes vanish and the internal space becomes Calabi-Yau. There is
however no direct limit to massless configurations related to
intersecting D6-branes, for which the torsion classes are: ${\cal W}_1
={\cal W}_3 =0$, but ${\cal W}_2, {\cal W}_4, {\cal W}_5\neq 0$
\cite{351}.

The differential equation of the spinor can be solved by a constant
spinor if one imposes first order differential equations on the
Vielbeine $e^n$
\be525
\textstyle{
\omega^{pq} J_{pq} = 0 \ , \quad 
\omega^{pq} \Omega_{pq}{}^n = - {\lambda \over 9} e^\phi e^n 
}
\ee
where $\omega^{pq} \equiv \omega^{pq}_m dy^m$ are the spin-connection
1-forms. Therefore, 6-d nearly K\"ahler spaces can be seen
as a weak SU(3)-holonomy space, which like Calabi-Yau spaces have, e.g.,
a vanishing first Chern class.  Their close relationship to special
holonomy spaces comes also due to the fact that the cone over nearly
K\"ahler 6-manifolds become a $G_2$-holonomy spaces \cite{690},
defined by a covariantly constant spinor. This can be verified by
multiplying (\ref{525}) with $J$ from the right and identifying the
rhs as the spin connection $\omega^{7n}$.  Note, the spin connection
1-form of $G_2$ holonomy spaces satisfy $\omega^{MN} \varphi_{MNP} =
0$, where $\varphi_{MNP}$ is the $G_2$-invariant 3-form. It is hence
straightforward to construct nearly K\"ahler spaces starting from
$G_2$ holonomy spaces and the almost K\"ahler form, that defines our
vacuum completely, is then given by $J_{mn} = \varphi_{mn7}$.  


\subsection*{Generic case}


If the complex coefficients $a$ and $b$ are generic, the solution
becomes more involved. Recall, in the cases discussed before either
the mass parameter had to vanish or the external space cannot be flat
and one might wonder whether this generic case allows for flat space
vacuum with non-zero mass parameter and therefore go beyond solutions
obtained from M-theory.

We need to consider only the singlets of the fluxes, because they are
related to the mass parameter or generate a non-zero cosmological
constant (or superpotential). If $F = F_0 J + \ldots$ and $G=G_0 \,
J\wedge J + \ldots$ and setting $W=0$, we find from (\ref{761}) that
$F_0=0$ and $m \sim G_0$. Then, the dilaton variation in (\ref{329})
gives, up to a real coefficient, the equation
\[
H \gamma^7 \chi = H (a \eta - b^\star \eta^\star) \ \sim \ m \chi
=m( a \eta + b^\star \eta^\star) \ .
\]
Using (\ref{230}), we find for singlet parts of $H$: $(H \haken
\Omega) \sim i {a \over b^\star}$ and $(H\haken \Omega^\star) \sim
- i{b^\star \over a}$. These equations are consistent only, if
\[
|a|^2 = |b|^2 \ .
\]
Hence, $a$ and $b$ differ only by a phase (ie.\ $a = r e^{i\alpha}$,
$b = r e^{i \beta}$) and we can write the spinor Ansatz
\[
\chi = a \eta + b^\star \eta^\star = r e^{i{\alpha - \beta \over 2}}
\Big[ \tilde \eta + \tilde \eta^\star \Big]
\]
with $\tilde \eta = e^{-i {\alpha + \beta \over 2}}\eta $. This
however, is equivalent to the Majorana case discussed before and
therefore, by investigating the internal variation, we will find
again, that non-trivial fluxes are only possible if the mass parameter
vanishes or if the cosmological constant is non-zero.


\section{Fluxes and geometry for SU(2) structures}

\resetcounter

In the case of SU(2) structures 
the internal spinor, entering (\ref{761}) -- (\ref{324}), becomes 
\[
\chi = a_i \eta_i + b^\star_i \eta_i^\star
\]
and for the sake of simplicity we will further drop the index $i$ and
keep in mind, that $a$ and $b$ are now vectors and $\eta$ is a spinor
doublet. Solutions with SU(2) structures are in general very involved,
and at this stage we shall not discuss the most general case. In
contrast with the unique solution that we found for SU(3) structures,
we are now interested in finding a specific flux vacuum whose 4-d
space-time is conformally Minkowski space and the internal space is
conformally flat.  Therefore, we will set in the following
\[
W = \lambda = 0 
\]
and write the metric as
\be553
ds^2 = e^{2A} [ -dt^2 + dx^i dx^i ]
+ e^{2B} [ dv^2 + du^2 + dz^1 d \bar z^1 + dz^2 d\bar z^2 ] 
\ee
where $\partial_u$ and $\partial_v$ are the two (global) vectors that
we introduced before. The internal metric becomes therefore:
$h_{mn}=e^{2(B-A)}\delta_{mn}$ and in the definition of $\FF$, $\GG$
and $\HH$ in (\ref{023}) we have to replace: $A \rightarrow B$.  The
existence of a flat vacuum with an internal Calabi-Yau space has been
suggested in \cite{120}, but it was unclear whether it can also be
made flat. As we will show now, there exist in fact such a vacuum and
the fields can be given explicitly. 

For our solution, the warp factors and the dilaton depend only on the
coordinates $u$ and $v$ so that
\[
A =  A(v,u) \quad , \qquad B = B(v,u) \quad , \qquad \phi = \phi(v,u) \ .
\]
With the (massive) 2-form $F$ also the $H$- and $G$-flux is fixed (we
assume here that $C_3=0$).  The component of $F$ proportional to $u
\wedge v$ does not contribute to the $H$-field and enter the equations
(\ref{761}) and (\ref{329}) in the same way as the Freud-Rubin
parameter $\lambda$ (but with different factors). By including this
component as well as the analogous component of $G$ (ie.\ where two
legs of the 4-form are along $u \wedge v$), we could not find other
solutions and therefore we set these flux components to zero from the
beginning.  Therefore, let us consider a 2-form restricted to the 4-d
base space spanned by the two complex coordinates $(z^1, z^2)$, which
allows in total for six 2-forms: three are the SU(2) singlets $J_0$,
Re$\hat \Omega$, Im$\hat \Omega$ and in real coordinates they can be
chosen to be selfdual. In addition there are three anti-selfdual
2-forms, which are, in respect to the symplectic form $J_0$, primitive
and of (1,1)-type. Each set, the selfdual and anti-selfdual forms obey
a quaternionic algebra and one finds that the 2-form components along
each of the three selfdual and three anti-selfdual are equivalent and
therefore we will pick just one selfdual and one anti-selfdual
component and write
\be524
F= m\,e^{2B}\, \Big[ f_0\,   {\rm Im} (dz^1 \wedge dz^2)
+ f_1{\rm Im} (dz^1 \wedge d \bar z^2) \Big]
\ee
where the constant $f_0$ parameterize the SU(2) singlet part and the
$f_1$ parameter is related to a primitive (1,1)-part of $F$.  Then, 3-
and 4-form (recall $C_3=0$) become
\[
\ba{rcl}
H &=& d e^{2B} \wedge \,\Big[ f_0 \,  {\rm Im} (dz^1 \wedge dz^2) 
+ f_1 {\rm Im} (dz^1 \wedge d\bar z^2 ) \Big]\ , \\[2mm]
G&=& {1 \over 2 m}\,  F \wedge F = m \, e^{4B}\, (f_0^2- f_1^2) \, vol_4 
\ea
\]
where $vol_4$ is volume form for the 4-d base base.  In the discussion
of the BPS eqs.\ (\ref{761}) -- (\ref{324}) we will start with the
last equation, which is the most constraining one.  Because $F$ and
$G$ have no components along the $(u,v)$ space, we obtain the
equations
\be552
\ba{l}
\Big[{1 \over 2} \gamma_u \gamma^v \partial_v(B-A) -
{1 \over 8} \HH_u \gamma^7 \Big] \chi = 0 \ , \\[2mm]
\Big[{1 \over 2} \gamma_v \gamma^u \partial_u(B-A) -
{1 \over 8} \HH_v \gamma^7 \Big] \chi = 0\ .
\ea
\ee
With the relations (\ref{541}) and $\HH_{npq} = {1 \over m}
\partial_n B \, \FF_{pq} +cycl$, one finds
\bea378
\HH_v \gamma^7 \chi &=& 4  f_0 \, \partial_v B \, [- ia\sigma_2 \sigma_3 
  \eta^\star +i b^\star \sigma_2 \sigma_3 \eta] \nonumber \ ,\\
\HH_u \gamma^7 \chi &=& 4  f_0 \, \partial_u B \, [-i a\sigma_2 \sigma_3 
\eta^\star +i
  b^\star \sigma_2 \sigma_3 \eta ] \nonumber \ , \\
\gamma_v \gamma^u \partial_u(B-A) \chi &=& \partial_u (B-A) \, 
[- i a \sigma_3 \eta + i b^\star \sigma_3 \eta^\star] \nonumber \ , \\
\gamma_u \gamma^v \partial_u(B-A) \chi &=& \partial_v (B-A) \, 
[ i a \sigma_3 \eta - i b^\star \sigma_3 \eta^\star] \nonumber \ .
\eea
Note, the $f_1$ part drops out here, because the SU(2) singlet spinor
$\chi$ picks up only the SU(2)-singlet component of the 2-form $F$,
which can be verified by employing the projectors in (\ref{230}).
With these expressions, the terms in (\ref{552}) cancel, if the
two complex vectors $a$ and $b^\star$ obey
\be771
b^\star = \pm  (a \sigma_2) 
\ee
[in the following we choose the ``+''-sign] and $A$ and $B$ have to
satisfy the equations
\be835
\ba{l}
 f_0 \partial_v B + \partial_u (B-A) = 0 \ , \\[2mm]
 f_0 \partial_u B - \partial_v (B-A) = 0 \ .
\ea
\ee
Since these are the Cauchy-Riemann equations, $B$ and $(B-A)$ can be
combined into one holomorphic function depending on the complex
coordinate $w = v + i\, u$. 

Using also the relation (\ref{947}) and $G_{rspq}\gamma^{rspq} = {1
\over 2m} [ (F_{rs} \gamma^{rs})(F_{pq} \gamma^{pq})+ 2 F_{pq}
F^{pq}]$, we find for the other terms in the BPS equations
(\ref{761}) and (\ref{329})
\be773
\ba{rcl}
\partial A \chi &=& 
\partial_v A \, [a \sigma_1 \eta - b^\star \sigma_1 \eta^\star] 
+ \partial_u A \, [a \sigma_2 \eta + b^\star \sigma_2 \eta^\star] 
\\[2mm]
\FF \gamma^7 \chi &=& 
4 m \,f_0 \,[ -i a\sigma_2 \sigma_3 \eta^\star + ib^\star \sigma_2 \sigma_3
  \eta ] 
\ ,\\[2mm]
\GG \chi &=& 4 m \, (f_1^2-f_0^2 )\, \chi  \ , \\[2mm]
\HH \gamma^7 \chi &=& - 12 f_0 \Big(
 \partial_v B \, [b^\star \eta + a  \eta^\star] 
+i \partial_u B \, [b^\star \sigma_3 \eta - a \sigma_3 \eta^\star] 
 \Big)\ .
\ea
\ee
Note, the $f_1$ parameter enters only the 4-form $G ={1 \over 2m} F
\wedge F$, which is proportional to the volume form and both terms of
$F$ enter with the opposite sign.  In the BPS equations these terms
can cancel only if we impose another constraint on $a$,
namely\footnote{A more general constraint like: $a = \cos\alpha
\, (\sigma_1 a) + \sin\alpha \, (\sigma_2 a)$ yields at the end the
same results.}
\be391
a = \sigma_1 a 
\ee
i.e.\ $a$ is an eigenvector to $\sigma_1$.  This fixes the vector $a$ and
due to (\ref{771}) also $b$, up an overall factor and the expression
in (\ref{773}) can be expressed in terms of the original spinor $\chi$
and we get
\be029
\ba{rcl}
\partial A \chi &=& 
[ \partial_v A  + i \gamma^7 \partial_u A ] \chi
\\[1mm]
\FF \gamma^7 \chi &=& 
4m f_0 i \gamma^7 \chi \ ,\\[1mm]
\GG \chi &=& 4 m \, (f_1^2 - f_0^2) \, \chi  \ , \\[1mm]
\HH \gamma^7 \chi &=&  12 f_0 [
\partial_u B + i\gamma^7 \partial_v B  ] \chi \ .
\ea
\ee
Therefore we find  from  (\ref{329})
\be886
\ba{rcl}
f_0 \partial_u B - \partial_v \phi &=& e^{\phi +B}
[\, {5\over 4}   + (f_1^2- f_0^2) ] \, m \ ,
\\[1mm]
f_0 \partial_v B + \partial_u \phi &=& {3 \over 4} e^{\phi +B}
f_0\, m
\ea
\ee
and in the external variation (\ref{761}), we use the Cauchy-Riemann
eqs.\ (\ref{835}) to replace $A$ by $B$ and find
\be813
\ba{rcl}
f_0 \partial_u B - \partial_v B &=&  e^{\phi +B}
[\, {1 \over 4}+ (f_1^2-f_0^2) \,] m
\\[1mm]
f_0 \partial_v B + \partial_u B &=& e^{\phi +B}
 m f_0
\ea
\ee
These two sets of equations fix the dilaton $\phi$ and $B$, which in
turn, due to (\ref{835}), fixes the warp factor $A$.

There is still one set of equations left, which are the
remaining components of the internal variation (\ref{324}) (we solved
so far only the $(u,v)$-part). If we contract this equation with
$\gamma^m$, we find
\be669
-5  \partial (B-A)\chi +{1 \over 4} H \gamma^7  \chi 
=  e^{\phi +B}
[ F \gamma^7 - 2 G] \chi \ .
\ee
Note, all functions are fixed and this equation gives only a
constraint.  Actually, inserting the expression (\ref{029}) one gets
two equations for the two parameter $f_0$ and $f_1$ and we found
\[
f_0^2 = {1 \over 8} \quad , \qquad f_1^2 = {3 \over 8} \ .
\]
Note, the mass parameter $m$ drops out on both side and is still a
free parameter.

We assumed here that the 2-form $F$ has one selfdual and one
anti-selfdual component only. In general one would write $F =e^{2B} [
f_i \omega^i + \tilde f_j \tilde \omega^j ]$, where $\omega^i$ are the
three self-dual 2-forms and $\tilde \omega^j$ are the three
anti-selfdual 2-forms. The calculation becomes more involved, but at
the end one gets the same equations with the only difference that:
$f_0 \rightarrow \pm \sqrt{{f_i f_i}}$ and $f_1 \rightarrow \pm
\sqrt{\tilde f_j \tilde f_j}$. This result might have been expected,
because the two sets of two forms satisfy a quaternionic algebra and
can be rotated into one another.

Finally, let us also discuss the equations of motion [since we
give the 2-form explicitly, the Bianchi identities are trivially
solved].  In the string frame they are given by
\[
d[ {}^\star G] =0 \quad , \qquad d [e^{-2\phi} {}^\star H]
= m \, F + 12\, F \haken G \ .
\]
Note, the rhs of the $G$-equation is zero because $G$ as well as $H$
have only internal components and since ${}^\star G \sim du \wedge
dv$, the 4-form equation is trivially solved for our choice of $A$ and
$B$.  On the other hand, in order to verify the $H$-field equation one
writes the lhs in components as
\[
{1 \over \sqrt{g}} \partial_m [
\sqrt{g} e^{-2\phi} H^{mpq} ] = e^{-4A-6B} \partial_m \Big[ e^{4 A
-2\phi} \partial_m e^{2B} \, \FF_{pq}]
\]
where $\FF_{pq}$ was the constant 2-form (without the factor $e^{2B}$)
and one uses our equations that imply that $B$ is harmonic and
$\partial_m (2 A - \phi +B) \partial_m B = {3\over 8} \, m^2 \,
e^{2(B+\phi)}$.


\section{Discussion of vacua with SU(3) structures}

\resetcounter


General SU(3) structures are very restrictive with respect to possible
fluxes. For example, for a non-zero mass parameter ($m\neq 0$) the
external space can be flat only if there are no fluxes.  For $m=0$,
there are two possibilities: either all RR-fields are zero and only
the 3-form $H$-flux is non-zero or only the RR-2-form is nonzero and
all other fluxes are trivial. Both cases are known flux vacua that can
be obtained from M-theory \cite{350, 120}. Our interest is in finding
vacua with SU(3) structures with non-zero mass \cite{780}. In this
case one is forced to add a cosmological constant (or superpotential).
As we have demonstrated, in this case the internal space has to be
nearly K\"ahler, which are Einstein and compact (for positive scalar
curvature).  In the following we shall discuss specific explicit
examples and their physics implications in more detail.


\subsection*{Fluxes and massive D2-brane}


The cone over any nearly K\"ahler manifold gives a 7-manifold with
$G_2$ holonomy \cite{690}, but due to the mass parameter, we
cannot identify the $7^{th}$ direction as the M-theory circle.  One
should instead identify this additional coordinate as the radial
direction of the (external) AdS space.  The 10-d metric can thus be
interpreted as the geometry of a massive D2-brane where the
transversal space has $G_2$ holonomy with non-zero fluxes.  In the
string frame, the metric becomes
\[
ds^2_{10} = {1 \over \sqrt{H(r)}} \Big[ -dt^2 + dx^2_1 + dx_2^2 \Big]
+ \sqrt{H(r)}\,  \Big[ dr^2 + r^2 d \Omega^{(NK)}_6 \Big]
\]
where $d \Omega^{(NK)}$ is the 6-dimensional nearly K\"ahler
metric. Our solution with four unbroken supercharges corresponds to
the case where $H \ \sim \ (m\, r)^{-4}$ with constant dilaton so that
the space factorizes into $AdS_4 \times \Omega^{(NK)}$. This might be
a specific limit of a more general solution, which either preserve
only two supercharges, i.e.\ gives an \N=1 vacuum in 3 dimensions, or
the complete solution breaks the SU(3) to SU(2)-structure. This is
very natural for (non-conical) $G_2$-holonomy spaces, because these
7-d spaces admit always for SU(2) structures \cite{800}. Examples of
such 2-brane solutions in the presence of fluxes are discussed in
\cite{CLP} or in M-theory in \cite{790}, which are however not
directly related to our solution, which has a non-zero mass parameter
and which has a fixed (constant) dilaton. It would be interesting to
explore this direction in more detail, but let us instead discuss some
examples of nearly K\"ahler spaces.

Starting with the corresponding $G_2$-holonomy space, one can obtain
explicit expressions for the metric and the almost complex structure
$J$ of the nearly K\"ahler 6-manifold; see \cite{720,730}.  There are
only a few known coset examples, which are discussed in more detail in
\cite{710, 760}.

$(i)\ {G_2 \over SU(3)} \simeq S_6$ \  This is a standard
example of a nearly K\"ahler space, where the cone becomes the flat
7-d space. Note, one can express the 6-sphere also by the coset
${SO(7)/SO(6)}$ which however breaks supersymmetry.

$(ii)\ {Sp(2) \over Sp(1) \times U(1)} \simeq {S_7 \over U(1)} \simeq
{\mathbb{CP}_3}$ \ The corresponding $G_2$-holonomy space is an
${\mathbb R_3}$ bundle over $S_4$ and hence it is the $SO(5)$
invariant metric of ${\mathbb{CP}_3}$ appearing here and not the
$SU(4)$-invariant, which is K\"ahler (instead of nearly
K\"ahler) and hence would break supersymmetry.

$(iii)\ {SU(3) \over U(1) \times U(1)}$ \ The cone over this space
gives the $G_2$-holonomy space related to an ${\mathbb R_3}$ bundle
over ${\mathbb{CP}_2}$ and therefore the 6-d metric is
$SU(3)$-invariant. This space is isomorphic to the flag manifold,
which again allows for another metric which is K\"ahler and 
would break supersymmetry.

$(iv)\ {SU(2)^3 \over SU(2)} \simeq S_3 \times S_3$ \ In constructing
this coset, there are different possibilities of modding out the
$SU(2)$ and the nearly K\"ahler space appearing in our context is
obtained by a diagonal embedding yielding as $G_2$ manifold an
${\mathbb R_4}$ bundle over $S_3$.

Actually there is a whole class of known non-homogeneous examples,
which are obtained from $G_2$ manifold given by an ${\mathbb R_3}$
bundle over {\em any} 4-d self-dual Einstein space, where the nearly
K\"ahler space becomes an $S_2$ bundle over the 4-d Einstein space,
which is also known as the twistor space; $(i)$ and $(ii)$ are just
the simplest (regular) examples in this class of solutions; other
examples can be found in \cite{750, 660}. The appearance of the 4-d
base manifold in these $G_2$ holonomy spaces also explains, why one
should expect only SU(2) structures for the complete (massive)
D2-brane solution mentioned at the beginning of this section.


\subsection*{Fluxes and intersecting branes}


An interesting question is whether this flux vacuum can support a
chiral gauge field theory. On the IIA side there are different
candidates giving rise to a 4-dimensional field theory: wrapped D4-,
D6- or D8-branes.  In addition to these D-branes one can also include
NS5-branes, which in the massive case have to be endpoints of
D6-branes or D4-branes.  But before we discuss NS5-branes we will
explore the situation where we have  added only D-branes in the 
background of the flux vacuum solution.

Consider first the field theory on (closed) D4-branes, which wrap a
1-cycle inside the nearly K\"ahler space. The construction of this
internal space via $G_2$-holonomy spaces excludes a non-contractible
$S^1$ and therefore, D4-branes can only appear in the internal space
as dipoles (or higher multipoles) or that the total D4-brane charge
vanishes.  In fact, due to non-zero RR-flux, the D4-branes will be
polarized as known from the Myers effect, where stability is ensured
by the non-trivial world volume fields \cite{820}. This implies
however, that the world volume spectrum cannot be chiral. The same
conclusion holds for D8-branes, that have to wrap a 5-cycle, again
contractible (as the dual 1-cycle), and hence only polarized D8-branes
can appear, which however do not give rise to a chiral spectrum.

\medskip

\noindent
{\em Wrapped intersecting  D6-branes}

\noindent
For D6-branes the situation is different. There are nearly K\"ahler
spaces with a non-trivial third homology class and hence there are
3-cycles upon which one can wrap D6-branes. For the coset examples
that we mentioned above, case $(iv)$ has the correct topology for
D6-branes to wrap around (different) 3-spheres. As discussed in
\cite{790} (see also \cite{760}), this space has three supersymmetric
3-cycles and if one wraps around each of them the same number of
D6-branes, they add up to zero in homology, so that the total D6-brane
charge is zero. Therefore, due to this geometric property of this
nearly K\"ahler space, there are {\it no orientifold planes} required
to cancel the D6-brane charge. This property is also reflected in
cohomology.  The D6-brane is charged under the RR-2-form, but since
the second cohomology is trivial for this space, the total charge has
to vanish. Note, the 2-form $F \sim J$ as well as the 3-form ${\rm Re}
\Omega$, which ``calibrates'' the supersymmetric 3-cycle, are not
closed, which is a crucial property of the nearly K\"ahler
spaces. These generalized calibrations have been discussed in more
detail in \cite{810} and the fact that the calibrating form is not
closed means that the volume is not minimized, but the action is
extremized \cite{890} [note, the RR-flux background induces on the
worldvolume a non-trivial interaction]. The fact that no orientifold
projections are required also implies that the geometry is not
deformed if one wraps the same number of D6-branes around each 3-cycle
-- it will only shift the mass parameter. It was moreover shown in
\cite{790} that the resulting spectrum on the common intersection of
all three stacks of branes wrapping the supersymmetric 3-cycle is
chiral. In fact, viewed from the tangent space, the stacks of
D6-branes intersect exactly in $120^{\rm o}$, giving rise to chiral
fermions \cite{130} at the intersection.

If we wrap $k$ D6-branes around each cycle, the resulting gauge group
would be $SU(k)^3$. To obtain other gauge groups one can consider
orbifolds of the original internal manifold, i.e.  the nearly K\"ahler
space $(iv)$ is replaced by $S^3/ \Z_m \otimes S^3/\Z_n$, i.e. one
wraps the D6-branes not around $S^3$ but the corresponding Lens space.
Note, these orbifolds do not change the local metric and will still
solve the equations, but the global structure is different and as a
consequence also the gauge group changes. In order to investigate this
in more detail, we have to analyze the wrapping of each stack of
D6-branes\footnote{We are grateful to Angel Uranga and Ralph
Blumenhagen for a discussion on this issue.}. One possibility for the
three supersymmetric 3-cycles is given by \cite{790}: $\{(1, 0),
(0,1), (-1,-1) \}$ where the two integers parameterize the third
homology of the space $S_3\otimes S_3$. In the case of the orbifold
there are now two effects. If the brane wraps the Lens space, say
$S_3/\Z_m$, and the orbifold acts freely, the rank of the gauge group
is reduced by a factor of $n$. On the other hand if the D6-branes are
fixed by the $\Z_n$ orbifold, one has to introduce additional
Chan-Paton factors and the group is split, ie.\ the gauge group
becomes $U(k)^n$ if we wrap $kmn$ D6-branes on $S_3/\Z_m$ and the
transversal space is $S_3/\Z_n$. The same happens for the other
6-branes wrapping the remaining supersymmetric 3-cycles.  Therefore,
if we wrap $(k\cdot m \cdot n)$ D6-branes on each of the\
supersymmetric 3-cycles of $S_3 / \Z_m \otimes S_3/ \Z_n$, the gauge
group will be: $U(k)^n\times U(k)^m \times U(k)^q$, where $q = mn/p$
and $p = gycd(mn)$. Note that the chiral matter, appearing at the
intersection of the two Lens spaces is in the bi-fundamental
representation of the corresponding gauge group factors.

\medskip

\noindent
{\em Wrapped NS5-branes}

\noindent
A chiral spectrum can also be obtained by allowing for NS5-branes
sources, i.e.\ $dH = n_5 \delta_{5}$ where $n_5$ is the number of
5-branes. As a consequence of the Bianchi identities of {\em massive}
supergravity, NS5-branes appear at the boundary of open D6-branes and
D4-branes.  For a non-compact internal space without fluxes, this
setup has been discussed by \cite{HZ}; let us explore its relevance
for our flux vacuum.  The chiral spectrum comes here from open strings
stretching between the D4- and D6-branes, where the 4-d field theory
of interest lives on the boundary of D4-branes and not on D6-branes.
Therefore the number of D4-branes is identified with the number of
colors $N_c$ and the number of D6-branes ($n_6$) gives the flavor
number. Recall, it is important to have a non-zero mass parameter $m$
which relates the number of branes to each other: $ n_6 = m \, n_5$.
To have only one chirality in the open string spectrum, it is
important that the D6-branes end on the NS5-branes. Otherwise, ie.\ if
the D6-branes are closed, open strings with both orientations would
appear \cite{HZ}.  In string theory, the mass parameter corresponds to
D8-branes and this relation can be understood due to creation of
D6-branes when the D8-brane passes through the NS5-brane.  So the
D6-brane ends also on a D8-brane.

In contrast to the case discussed before, we need now non-contractible
supersymmetric 2-cycles in order to wrap NS5-branes, i.e.\ the second
Betti number should be non-zero $b_2 \neq 0$. Having these NS5-branes
and the mass parameter the D6- and D4-branes are open with the
endpoints fixed by the NS5-branes. Therefore, we should not consider
the example $(iv)$ from our list of nearly K\"ahler space, but instead
consider the case $(ii)$ or $(iii)$. The example of $\C \PP_3$ might
be too trivial, since it has only one supersymmetric 2-cycle ($b_2=1$)
to wrap the NS5-brane. More interesting might be the flag manifold:
${SU(3) \over U(1) \times U(1)}$ which has $b_2 = 2$ and, as in 
 \cite{HZ}, we can wrap two NS5-branes between which D4-branes
are stretched and which fix the endpoints of D6-branes. Thus the D4-
and D6-branes are stable even without any proper supersymmetric cycle
in the internal space.  In contrast to the case before, the resulting
gauge group is fixed only by the number of D4-branes, which connect
the two stacks of NS5-branes; the D6-branes also ending on at
NS5-branes give two flavor group factors.  Of course, one can also
setup more complicated configurations by using spaces with $b_2 > 2$,
e.g.\ by building twistor spaces over the self-dual Einstein space
found in \cite{830}.

A subtle question is the mass parameter, which is a crucial ingredient
of this vacuum solution. If one keeps $m$ just as a parameter and does
not introduce D8-branes, the open D6-branes have to end on both sides
on the NS5-branes. This however does not yield a chiral spectrum given
by 4-6 strings, because for this it is important that the open
D6-branes terminate on the NS5-brane only from one side \cite{HZ}.  A
better setup is given, if there are polarized or dipole D8-branes, so
that the total D8-brane charge is zero [recall, due to the absence of
supersymmetric 5-cycle we cannot simply wrap D8-brane on this space]
and the D6-branes can stretch between the NS5- and D8-branes.

\medskip

\noindent
{\em Moduli fixing}

\noindent
A phenomenological viable model requires that all moduli are fixed.
This is especially important in a cosmological setting, because
massless scalars can absorb a non-zero vacuum energy.  One has to
distinguish between closed and open string moduli, where the latter
ones are related to the brane position, i.e.\ they are fixed if the
brane wraps a rigid cycle. The closed string moduli on the other hand
correspond to deformations of the geometry. This discussion is very
difficult, because the corresponding moduli spaces are not yet well
understood in general.  But for the simple cases that we discussed in
more detail, the moduli space is finite and a discussion is given in
\cite{810}.

We expect that the closed string moduli are fixed by the (quantized)
fluxes that cover all non-trivial cycles of the geometry. Note, due to
the back-reaction of fluxes, the geometry of the internal space had to
change: from the original flat or Calabi Yau (with no fluxes) to the nearly
K\"ahler one. In the moduli space, this change in geometry is reflected
in a lifting of moduli.

But the moduli space is not completely lifted and the remaining moduli
should be related to open string moduli.  E.g.\ the $S^3 \times S^3$
geometry has a 3-dimensional moduli space \cite{810}, which, if we
wrap branes on the supersymmetric 3-cycles, is related to moving
around these branes relative to each other. So, although the three
supersymmetric 3-cycles are (generalized) calibrated, their relative
position to each other is not fixed. We expect something analogous
also for the second case, where NS5-branes wrap supersymmetric
2-cycles. The appearance of moduli on the world-volume of the
NS5-brane triggers in addition the non-rigidity of the open D4-brane
that ends on the two NS5-branes, as well as non-rigid open D6-branes
that on one end attach to the NS5-branes, and on the other end attach
to the D8-branes which are also not rigid.


\subsection*{Relaxation of the cosmological constant}


A negative four-dimensional cosmological constant is a generic feature
of supersymmetric flux vacua. In fact, for the massive Type IIA string
theory, the internal space with SU(3) structures and a non-zero mass
parameter impose that the four-dimensional space to be anti-deSitter.
In the 10-d Einstein frame, the negative cosmological constant was
given by
\[
\textstyle{
\Lambda = - e^{4\phi} \, m^2 = - \big( \, {H_0 \over 8 G_0} \, \big)^4 m^2
}
\]
where we inserted the solution for the fixed dilaton in terms of the
ratio of the quantized fluxes.  Recall, the fluxes $H_0$ and $G_0$
were the SU(3) singlet components of the 3-form and 4-form flux
introduced in (\ref{774}); they should be quantized if we take into
account the conditions of charge quantization of the corresponding
branes. By choosing flux vacua where the appropriate fluxes satisfy
($H_0 \ll G_0$), one can lower the quantized cosmological constant,
which however always remains negative.

The goal, however, is to obtain a small positive cosmological
constant. Within our framework we shall comment here on two possible
scenarios: $(i)$ by using the Brown-Teitelboim mechanism of
neutralization by membrane instantons \cite{850} or $(ii)$ to consider
a similar setup as KKLT scenario on the Type IIB side \cite{840}.  The
crucial ingredient in the latter approach are anti-branes, which
increase the vacuum energy and are meta-stable if there are no
appropriate fluxes.  In the KKLT setup one has anti-D3-branes plus
3-form fluxes and the anti-D3-branes can only decay if they blow up
into anti D5/NS5-branes (due to the Myers effect) which decay with the
3-form flux (bubble decay of the vacuum).  A small positive
cosmological constant can then be obtained by fine tuning. Note, that
it is important in this setup that there are no fluxes under which the
(anti) branes are charged; otherwise they can immediately decay, i.e.\ a
metastable minimum would not occur. This is exactly what we expect if
we apply this scenario to our flux vacuum, where all RR- and
NS-NS-fluxes are non-zero. Thus, whether we wrap anti-D6- or
anti-NS5-branes, we will not obtain a meta-stable deSitter minimum.

On the other hand, the Brown-Teitelboim mechanism $(i)$ is much more
appropriate for our supersymmetric vacuum solution.  This setup was
introduced in string theory in \cite{880,860,870} and let us summarize
some basic features. The crucial ingredients here are membrane
instantons, which relax the cosmological constant in the external
space. This is analogous to the decay of an electric field by
Schwinger pair creation of charged particles.  In Type IIA
supergravity we have D2-branes, which appear in the 4-d external space
as domain walls and separates regions with a different cosmological
constant, where the jump in the cosmological constant is proportional
to the brane charge. One does not need to consider strictly D2-branes,
also a wrapped brane, which extends in two spatial directions in the
external space appears as domain wall across which the cosmological
constant jumps.  As shown in the above literature in more detail,
these instantons can lift the negative cosmological constant to a
positive one and subsequently reach a deSitter vacuum.  Due to false
vacuum decay, this vacuum will not be stable, the most likely
configuration is the one with the smallest cosmological constant
\cite{870}. In order to argue for a very small positive value, one
needs an additional input. First, the decay from a positive to a
negative cosmological constant is gravitationally suppressed
\cite{900}, but in order to come sufficiently close to a vanishing
cosmological constant, one needs a small spacing in the jumps. Bousso
and Polchinski \cite{860} argued that a dense discretuum is generated
due to multiple flux possibilities appearing in compactification of
string or M-theory, but this is not enough to explain the observed
smallness. Instead, due to a (weak) anthropic selection they end up with
a vacuum of small positive cosmological constant.  On the other hand,
the implementation of the Brown-Teitelboim mechanism in \cite{870},
relies especially on two additional inputs: the brane may wrap an
internal degenerate cycle and the world volume dynamics give rise to
large density of state factor. Both effects yield a dynamical
relaxation of the cosmological constant to even smaller values.

It is straightforward to apply these arguments to our case.  One way
is to consider D2-branes to relax the cosmological constant and if the
NS-flux is sufficiently small, also the dilaton becomes small ($e^\phi
\ll 1$) yielding a small spacing in the discrete jumps.  But one can
also consider other branes, e.g.\ the D8-branes in our second example
(with wrapped NS5-branes), can also appear as domain walls in the
external space if they wrap the whole internal space and as in
\cite{870} this might yield an efficient relaxation of the
cosmological constant. Note however, both scenarios are non-BPS,
because the 4-d external space is parallel to a D4-brane and a
D2-brane as domain wall is non-BPS configuration. The same holds
also for D8-branes, which are non-BPS if they are non-parallel to
D4-branes (also D6-branes would not be BPS if they appear as domain
walls in the external space). So, whatever brane relaxes the 4-d
cosmological constant, it will break supersymmetry.


\bigskip

\bigskip


\section*{Acknowledgments}
We would like to thank Ralph Blumenhagen, Gianguido Dall'Agata,
Sergei Gukov, Ruben
Minasian, Gao Peng and Angel Uranga for discussions.  We are grateful
to the University of Pennsylvania (K.B.), the CERN Theory Division
(M.C.), the Albert-Einstein Institute (M.C.) for hospitality and
support during the course of this work.  Research is supported in part
by DOE grant DOE-EY-76-02-3071 (M.C.), NSF grant INT02-03585 (M.C.)
and Fay R. and Eugene L. Langberg endowed Chair (M.C.).


\bibliography{0407263}        
\bibliographystyle{utphys}   


\end{document}